\documentclass[10pt,final,journal,letterpaper,twoside,twocolumn]{IEEEtran}

\usepackage[cmex10]{amsmath}
\usepackage{amsfonts}
\usepackage{amssymb}
\usepackage[dvips]{graphicx}
\usepackage{dsfont}
\usepackage{psfrag}
\usepackage[tight,footnotesize]{subfigure}
\usepackage{cite}
\usepackage{balance}
\DeclareMathOperator{\sgn}{sgn}
\newtheorem{theorem}{Theorem}
\newtheorem{lemma}{Lemma}
\newtheorem{corollary}{Corollary}

\begin{document}

\title{On the Monotonicity of the Generalized Marcum and Nuttall \emph{Q}-Functions\authorrefmark{2}}

\author{Vasilios~M.~Kapinas,~\IEEEmembership{Member,~IEEE,}
Sotirios~K.~Mihos,~\IEEEmembership{Student~Member,~IEEE,}
and~George~K.~Karagiannidis,~\IEEEmembership{Senior~Member,~IEEE}%
\thanks{Manuscript received December 16, 2007; revised January 10,
2009. This work was supported in part by the Satellite
Communications Network of Excellence (SatNex) project (IST-507052)
and its Phase-II, SatNex-II (IST-27393), funded by the European
Commission (EC) under its FP6 program. The material in this paper
was presented in part at the International Symposium on Wireless
Pervasive Computing 2008,
Santorini, Greece, May 2008.}%
\thanks{The authors are with the Electrical and Computer Engineering
Department, Aristotle University of Thessaloniki, 54124
Thessaloniki, Greece (e-mail: kapinas@auth.gr; sotmihos@gmail.com;
geokarag@auth.gr).\par}
\thanks{\authorrefmark{2}This work is dedicated to the memory of Marvin K. Simon.}
}

\markboth{Published in IEEE Transactions on Information Theory,~Vol.~55,
No.~8,~August~2009}{Kapinas \MakeLowercase{\textit{et al.}}: On the
Monotonicity of the Generalized Marcum and Nuttall
\emph{Q}-Functions}


\maketitle

\begin{abstract}
\boldmath Monotonicity criteria are established for the generalized
Marcum \emph{Q}-function, $\emph{Q}_{M}(\alpha,\beta)$, the standard
Nuttall \emph{Q}-function, $\emph{Q}_{M,N}(\alpha,\beta)$, and the
normalized Nuttall \emph{Q}-function,
$\mathcal{Q}_{M,N}(\alpha,\beta)$, with respect to their real order
indices $M,N$. Besides, closed-form expressions are derived for the
computation of the standard and normalized Nuttall
\emph{Q}-functions for the case when $M,N$ are odd multiples of
$0.5$ and $M\geq N$. By exploiting these results, novel upper and
lower bounds for $\emph{Q}_{M,N}(\alpha,\beta)$ and
$\mathcal{Q}_{M,N}(\alpha,\beta)$ are proposed. Furthermore,
specific tight upper and lower bounds for
$\emph{Q}_{M}(\alpha,\beta)$, previously reported in the literature,
are extended for real values of $M$. The offered theoretical results
can be efficiently applied in the study of digital communications
over fading channels, in the information-theoretic analysis of
multiple-input multiple-output systems and in the description of
stochastic processes in probability theory, among others.
\end{abstract}


\section{Introduction}

\subsection{The Nuttall Q-Functions}\label{subsec:Nuttall}

\IEEEPARstart{A}{n} extended version of the (standard) Marcum
$\emph{Q}$-function, $\emph{Q}(\alpha,\beta)=\int_{\beta}^{\infty}x
e^{-\frac{x^2+\alpha^2}{2}} I_{0}(\alpha x) dx$, where
$\alpha,\beta\geq 0$, originally appeared in
\cite[Appendix,~eq.~(16)]{Marcum_IRE_statistical}, defines the
standard\footnote{We adopt the term \textquotedblleft
standard\textquotedblright \ for the Nuttall $\emph{Q}$-function in
order to avoid ambiguity with its normalized version to be
introduced later.} Nuttall \emph{Q}-function
\cite[eq.~(86)]{Nuttall_NUSC_some_integrals}, given by the integral
representation
\begin{align}\label{eq:st_Nuttall_definition}
\emph{Q}_{M,N}(\alpha,\beta)=\int_{\beta}^{\infty}x^M
e^{-\frac{x^2+\alpha^2}{2}} I_{N}(\alpha x) dx
\end{align}
where the order indices are generally reals with values $M\geq 0$
and $N>-1$, $I_{N}$ is the $N$th order modified Bessel function of
the first kind \cite[eq.~(9.6.3)]{Abramowitz_book} and
$\alpha,\beta$ are real parameters with $\alpha>0$, $\beta \geq 0$.
It is worth mentioning here, that the negative values of $N$,
defined above, have not been of interest in any practical
applications so far. However, the extension of the Nuttall
\emph{Q}-function to negative values of $N$ has been introduced here
in order to facilitate more effectively the relation of this
function to the more common generalized Marcum \emph{Q}-function, as
will be shown in the sequel. An alternative version of
$\emph{Q}_{M,N}(\alpha,\beta)$ is the \emph{normalized Nuttall
Q-function}, $\mathcal{Q}_{M,N}(\alpha,\beta)$, which constitutes a
normalization of the former with respect to the parameter $\alpha$,
defined simply by the relation
\begin{align}\label{eq:norm_Nuttall_definition}
\mathcal{Q}_{M,N}(\alpha,\beta) \triangleq
\frac{\emph{Q}_{M,N}(\alpha,\beta)}{\alpha^N}.
\end{align}

Typical applications involving the standard and normalized Nuttall
\emph{Q}-functions include: (a) the error probability performance of
noncoherent digital communication over Nakagami fading channels with
interference \cite{Simon_Nuttall}, (b) the outage probability of
wireless communication systems where the Nakagami/Rician faded
desired signals are subject to independent and identically
distributed (i.i.d.) Rician/Nakagami faded interferers,
respectively, under the assumptions of minimum interference and
signal power constraints
\cite{Simon_Nuttall,Yang_&_Alouini,Yang_On_the_average,Jerez_conf_antenna},
(c) the performance analysis and capacity statistics of uncoded
multiple-input multiple-output (MIMO) systems operating over Rician
fading channels \cite{Kang_eigenvalue,Jin_beamforming,Aissa}, and
(d) the extraction of the required log-likelihood ratio for the
decoding of differential phase-shift keying (DPSK) signals employing
turbo or low-density parity-check (LDPC) codes \cite{Ho_conf_soft}.

Since both types of the Nuttall $\emph{Q}$-function are not
considered to be tabulated functions, their computation involved in
the aforementioned applications was handled considering the two
distinct cases of $M+N$ being either odd or even, in order to
express them in terms of more common functions. The possibility of
doing such when $M+N$ is odd was suggested in
\cite{Nuttall_NUSC_some_integrals}, requiring particular combination
of the two recursive relations \cite[eqs.~(87),
(88)]{Nuttall_NUSC_some_integrals}. However, the explicit solution
was derived only in \cite[eq.~(13)]{Simon_Nuttall} entirely in terms
of the Marcum $\emph{Q}$-function and a finite weighted sum of
modified Bessel functions of the first kind. Having all the above in
mind, along with the fact that the calculation of
$\emph{Q}(\alpha,\beta)$ itself requires numerical integration, the
issue of the efficient computation of
\eqref{eq:st_Nuttall_definition} and
\eqref{eq:norm_Nuttall_definition} still remains open.

\subsection{The Generalized Marcum Q-Function}\label{subsec:Marcum}
\IEEEpubidadjcol The generalized Marcum \emph{Q}-function
\cite{Marcum_RAND_Table} of positive real order $M$, is defined by
the integral~\cite[eq.~(1)]{Nuttall_IEEE_some_integrals}
\begin{align} \label{eq:Marcum_definition}
\emph{Q}_{M}(\alpha,\beta) \triangleq \frac{1}{{\alpha}^{M-1}} \int_{\beta}^{%
\infty} x^M e^{-\frac{x^2+{\alpha}^2}{2}}I_{M-1}(\alpha x)dx
\end{align}
where $\alpha$, $\beta$ are non-negative real
parameters\footnote{For $\alpha=0$ the right hand side of
\eqref{eq:Marcum_definition} can be easily shown to satisfy the
limiting value of \cite[eq.~(4.71)]{Simon_Fading_channels_book},
reproduced in \eqref{eq:Sotos_Marcum_zero}.}. For $M=1$, it reduces
to the popular standard (or first-order) Marcum \emph{Q}-function,
$\emph{Q}_1(\alpha,\beta)$ (or $\emph{Q}(\alpha,\beta)$), while for
general $M$ it is related to the normalized Nuttall
\emph{Q}-function according to
\cite[eq.~(4.105)]{Simon_Fading_channels_book}
\begin{align}\label{eq:Marcum-Nuttall_relation}
\emph{Q}_{M}(\alpha,\beta)=\mathcal{Q}_{M,M-1}(\alpha,\beta),\quad
\alpha>0.
\end{align}

An identical function to the generalized Marcum \emph{Q} is the
probability of detection\footnote{For $M$ incoherently integrated
signals, the two functions are simply related by
$\emph{Q}_{M}(\alpha,\beta)=P_{M}(\frac{\alpha^2}{2M},\frac{\beta^2}{2})$,
as induced by \cite[eq.~(7)]{Shnidman_The_calculation}.}
\cite[eq.~(49)]{Marcum_IRE_statistical}, which has a long history in
radar communications and particularly in the study of target
detection by pulsed radar with single or multiple observations
\cite{Marcum_IRE_statistical,Swerling_Probability,Helstrom_book,Trees_Detection_book}.
Additionally, $\emph{Q}_{M}(\alpha,\beta)$ is strongly associated
with: (a) the error probability performance of noncoherent and
differentially coherent modulations over generalized fading channels
\cite{Simon_A_unified_approach,Simon_A_unified_performance,Simon_Some_new_results,
Simon_Fading_channels_book,Gaur_some_integrals,Annamalai_error_prob},
(b) the signal energy detection of a primary user over a multipath
channel \cite{Simon_energy,Pandharipande_conf_performance}, and
finally (c) the information-theoretic study of MIMO systems
\cite{Zhu_On_the_mutual}. Aside from these applications, the
generalized Marcum \emph{Q}-function presents a variety of
interesting probabilistic interpretations. Most indicatively, for
integer $M$, it is the complementary cumulative distribution
function (CCDF) of a noncentral chi-square ($\chi^2$) random
variable with $2M$ degrees of freedom (DOF)
\cite[eq.~(2.45)]{Simon_Gaussian_book}. This relationship was
extended in \cite{Shnidman_two_noncentral} to work for the case of
odd DOF as well, through a generalization of the noncentral $\chi^2$
CCDF. Similar relations can be found in the literature involving the
generalized Rician \cite[(2.1--145)]{Proakis_book}, the generalized
Rayleigh \cite[pp.~1]{Omura_Stanford_Numerical} (for $\alpha=0$) and
the bivariate Rayleigh \cite[Appendix~A]{Schwartz_book},
\cite{Simon_A_simple_single} (for $M=1$) CCDF's. Finally, in a
recent work \cite{Weinberg}, a new association has been derived
between the generalized Marcum \emph{Q}-function and a probabilistic
comparison of two independent Poisson random variables.

More than thirty algorithms have been proposed in the literature for
the numerical computation of the standard and generalized Marcum
\emph{Q}-functions, among them power series expansions
\cite{Robertson_chi-square,Dillard,Shnidman_Efficient_evaluation},
approximations and asymptotic expressions
\cite{Rappaport,Helstrom_computing,Weinberg_DSTO_Numerical,Ding_Marcum},
and Neumann series expansions
\cite{Johansen_ARL_New_techniques,Parl,Cantrell}. However, the above
representations may not always provide sufficient information about
the relative position of the approximated value with respect to the
exact one, which in some applications is highly desired. In
\cite{Simon_new_twist}, the generalized Marcum \emph{Q}-function of
integer order $M$ has been expressed as a single integral with
finite limits, which is computationally more desirable relatively to
other methods suggested previously. Nevertheless, the integral
cannot be computed analytically and appropriate numerical
integration techniques have to be applied, thereby introducing an
approximation error in its computation. In \cite{Ross}, an exact
representation for $\emph{Q}_{M}(\alpha,\beta)$, when $M$ is an odd
multiple of $0.5$, was given as a finite sum of tabulated functions,
assuming that $\beta^2>\alpha^2+2M$. This result was recently
enhanced in \cite{Li_conf_computing} to a single expression that
remains accurate over all ranges of the parameters $\alpha,\beta$,
while in \cite{Giuseppe_Jenses-Cotes} the same expression was
bounded by particular utilization of novel Gaussian
\emph{Q}-function inequalities. Finally, in \cite{Sun+conf_tight},
an equivalent expression to \cite[eq.~(11)]{Li_conf_computing} was
derived, adopting a completely different (analytical) approach from
the latter.

Close inspection of the issues mentioned above, render the existence
of upper and lower bounds a matter of essential importance in the
computation of \eqref{eq:Marcum_definition}. Several types of bounds
for the standard
\cite{Corazza_Marcum,Kam&Li_comp_and_bound,Zhao_Tight_geometric} and
generalized
\cite{Rappaport,Li_conf_computing,Li_conf_generic,Simon_Exponential,
Annamalai,Sun+conf_tight} Marcum \emph{Q}-functions have been
suggested so far. However, all the aforementioned works consider
just integer values of $M$, which is generally true when this
parameter represents the number of independent samples of a
square-law detector output. Nevertheless, in many applications, this
requirement does not hold. According to
\cite[Sec.~4.4.2]{Simon_Fading_channels_book}, it would be desirable
to obtain alternative representations for
$\emph{Q}_{M}(\alpha,\beta)$ regardless of whether $M$ is integer or
not. For instance, in \cite{Simon_A_unified_performance}, the fading
parameter of the Nakagami-$m$ distribution is restricted to integer
values in the lack of a closed-form expression for the generalized
Marcum \emph{Q}-function of real order. Additionally, in
\cite{Mills,Helstrom_Inversion,Pandharipande_conf_performance,Herath_conf_analysis},
the order $M$ of the generalized Marcum \emph{Q}-function, involved
in the energy detection in various radiometer and cognitive radio
applications, is expressed as the product of the integration time
and the receiver bandwidth, thus implying that in general $M$ is a
non-integer quantity. Furthermore, a probabilistic interpretation of
$\emph{Q}_{M-\mu}(\alpha,\beta)$, where $M\in
\mathds{N}$\footnote{Throughout the manuscript, we adopt
$\mathds{N}$ and $\mathds{N}_0$ notations for the representation of
the positive and the non-negative integer set, respectively. In the
same way, $\mathds{R}^+$ includes the positive and $\mathds{R}^+_0$
the non-negative reals.} and $\mu=0.5$, is given in
\cite{Di_Blasio}, where it is related to main probabilistic
characteristics of $2(M-\mu)$ random variables, while in
\cite{Patnaik_non-central,Ghosh_monotonicity,Temme_Asymptotic},
noncentral $\chi^2$ random variables with fractional DOF are
studied. Finally, in \cite{Baricz_inequalities}, the integrand of
\eqref{eq:Marcum_definition} has been proved to be a probability
density function (PDF) for $\alpha\geq 0$ and $M>0$, a result that
also has been utilized in \cite{Sun+conf_tight}.

\subsection{Contribution}

As described in Subsection~\ref{subsec:Nuttall}, a closed-form
expression for the computation of the standard and normalized
Nuttall \emph{Q}-functions is available in the literature only for
the case of odd $M+N$, with the additional restriction of integers
$M,N$. In Subsection~\ref{subsec:Nuttall_closed}, we derive a novel
closed-form expression for the computation of
$\emph{Q}_{M,N}(\alpha,\beta)$ and $\mathcal{Q}_{M,N}(\alpha,\beta)$
when $M,N$ are odd multiples of $0.5$ and $M \geq N$, being valid
for all ranges of the parameters $\alpha,\beta$.

Besides, in Subsection~\ref{subsec:Nuttall_bounds}, we proceed with
the establishment of appropriate monotonicity criteria, revealing the
behavior of both functions with the sum $M+N$. Specifically, we
demonstrate that the standard Nuttall \emph{Q}-function is strictly
increasing with respect to $M+N$ when $M\geq N+1$, under the
constraints of $\alpha\geq 1$ and $\beta>0$. For the normalized
Nuttall \emph{Q}-function, a similar monotonicity statement is
proved without the necessity of reducing the range of $\alpha$.

An alternative approach, sufficient enough to facilitate the problem
of evaluating the Nuttall \emph{Q}-functions, is the derivation of
tight bounds. Nevertheless, to the best of the authors' knowledge,
such bounds have not been reported in the literature so far.
Subsection~\ref{subsec:Nuttall_bounds} is completed with the
exploitation of the previous results in order to derive novel upper
and lower bounds for $\emph{Q}_{M,N}(\alpha,\beta)$ and
$\mathcal{Q}_{M,N}(\alpha,\beta)$ when $M \geq N+1$ and $\beta>0$,
with the extra requirement of $\alpha\geq 1$ for the former.

Additionally, in Subsection~\ref{subsec:Marcum}, the need for
computing the generalized Marcum \emph{Q}-function,
$\emph{Q}_{M}(\alpha,\beta)$, of real order $M$ was highlighted,
since it is a case of frequent occurrence in various applications.
However, a thorough literature search for studies concerning
arbitrary values of $M$, revealed only \cite{Li_conf_computing} for
the closed-form computation of $\emph{Q}_{M}(\alpha,\beta)$ of
half-odd integer order, and the accepted paper
\cite{Baricz_Marcum_I}, where bounds for
$\emph{Q}_{M}(\alpha,\beta)$ were introduced for the case when $M$
is not necessarily an integer. These considerations motivated us to
generalize the scope of $M$ in \cite[eq.~(16)]{Li_conf_computing}
\begin{align}\label{eq:Kam_Marcum_inequality}
\emph{Q}_{M-0.5}(\alpha,\beta)<\emph{Q}_{M}(\alpha,\beta)<\emph{Q}_{M+0.5}(\alpha,\beta),\quad M \in \mathds{N}
\end{align}
as described in Section~\ref{sec:Marcum_monotonicity}, by providing
a monotonicity formalization for the generalized Marcum
\emph{Q}-function, namely that $\emph{Q}_{M}(\alpha,\beta)$ is
strictly increasing with respect to its order $M>0$ for $\alpha\geq
0$ and $\beta>0$. This interesting statement was also recently
presented in \cite{Baricz_inequalities}, using a different approach.
As a consequence, novel upper and lower bounds for
$\emph{Q}_{M}(\alpha,\beta)$ of positive real order are derived. We
finalize the paper with some concluding remarks, given in
Section~\ref{sec:Conclusion}.

\section{Monotonicity of the Nuttall \emph{Q}-Functions}\label{sec:Nuttall_monotonicity}

\subsection{Novel Closed-Form Representations}\label{subsec:Nuttall_closed}

So far, closed-form expression for either type of the Nuttall
\emph{Q}-function is not available in the literature. In this
section, we derive such a representation for the case when $M,N$ are
odd multiples of $0.5$ and $M \geq N$, through the theorem and
corollary established below. Before proceeding further with the
corresponding proofs, some definitions of essential functions and
notations used, would be very convenient.

Hereafter, $\Gamma$, $\gamma$ and $\Gamma \left(\cdot,\cdot\right)$
will denote the Euler gamma \cite[eq.~(6.1.1)]{Abramowitz_book}, the
lower incomplete gamma \cite[eq.~(6.5.2)]{Abramowitz_book} and the
upper incomplete gamma \cite[eq.~(6.5.3)]{Abramowitz_book}
functions, respectively, defined by the integrals
\begin{align*}
\Gamma(z)&=\int_0^{\infty}t^{z-1}e^{-t}dt, \
\gamma\left(z,x\right)=\int_0^{x}t^{z-1}e^{-t}dt\\
&\Gamma\left(z,x \right)=\Gamma(z)-\gamma\left(z,x\right), \quad
z\in\mathds{R}^+,x\in\mathds{R}.
\end{align*}
Notations $n!$, $(m)_n$ and $\binom {m}{n}$ imply the factorial
\cite[eq.~(6.1.6)]{Abramowitz_book}, the rising factorial
(Pochhammer's symbol) \cite[eq.~(6.1.22)]{Abramowitz_book} and the
binomial coefficient \cite[eq.~(24.1.1 C)]{Abramowitz_book},
respectively, defined by $n!=\prod_{k=1}^n k$ for $n\in\mathds{N}$;
$=1$ for $n=0$, $(m)_n=\frac{(m+n-1)!}{(m-1)!}$ for
$m\in\mathds{N}$, $n\in\mathds{N}_0$ and $\binom
{m}{n}=\frac{m!}{n!(m-n)!}$ for $m,n\in\mathds{N}_0$, $m\geq n$.
Finally, $\sgn(z)=z/|z|$ for $z\neq 0$; $=0$ for $z=0$, stands for
the signum function.
\begin{theorem}[Closed-form for the standard Nuttall Q]\label{theorem:st_Nuttall_exact}
The standard Nuttall \emph{Q}-function, $\emph{Q}_{M,N}(\alpha,\beta
)$, when $m=M+0.5 \in \mathds{N}$, $n=N+0.5 \in \mathds{N}$ and
$M\geq N$, can be evaluated for $\alpha>0$, $\beta\geq 0$ by the
following closed-form expression:
\begin{align*}
\emph{Q}_{M,N}(\alpha,\beta)={}&\frac{(-1)^n
(2\alpha)^{-n+\frac{1}{2}}}{\sqrt{\pi}}\\
&\times\sum_{k=0}^{n-1}\frac{(n-k)_{n-1}(2\alpha)^k}{k!}
\mathcal{I}_{m,n}^{k}(\alpha,\beta)
\end{align*}
where the term $\mathcal{I}_{m,n}^{k}(\alpha,\beta)$ is given by
\begin{align}
\mathcal{I}_{m,n}^{k}(\alpha,\beta)={}&(-1)^{k+1}\sum_{l=0}^{m-n+k}
\binom{m-n+k}{l}2^{\frac{l-1}{2}}\alpha^{m-n+k-l}\nonumber\\
&\times\left[(-1)^{m-n-l-1}\Gamma{\left(\frac{l+1}{2},
\frac{(\beta+\alpha)^2}{2}\right)}\right.\nonumber\\
&\left.-\left(\sgn(\beta-\alpha)\right)^{l+1}\gamma\left(\frac{l+1}{2},\frac{(\beta-\alpha)^2}{2}
\right)\right.\nonumber\\
&\left.+\Gamma\left(\frac{l+1}{2}\right)\right].\label{eq:Sotos_integral0}
\end{align}
\end{theorem}
\begin{IEEEproof}
Given that $n=N+0.5 \in \mathds{N}$, the modified Bessel function of
the first kind, $I_N$, can be expressed by the finite sum
\cite[eq.~(8.467)]{Gradshteyn_book}, which after some manipulations
can be written as
\begin{align}\label{eq:Sotos_mod_Bessel}
I_{N}(z)={}&\frac{(-1)^n (2z)^{-n+\frac{1}{2}}}{\sqrt{\pi} e^z}
\sum_{k=0}^{n-1} \frac{(n-k)_{n-1}(2z)^k}{k!}\nonumber\\
&\times\left(1-(-1)^k e^{2z} \right), \quad
n=N+\frac{1}{2}\in\mathds{N},z\in\mathds{R}.
\end{align}
Therefore, using \eqref{eq:st_Nuttall_definition} and
\eqref{eq:Sotos_mod_Bessel}, the standard Nuttall \emph{Q}-function
satisfies
\begin{align}\label{eq:Integral_difference}
\emph{Q}_{M,N}(\alpha,\beta)={}&\frac{(-1)^n
(2\alpha)^{-n+\frac{1}{2}}}{\sqrt{\pi}} \sum_{k=0}^{n-1}
\frac{(n-k)_{n-1}(2\alpha)^k}{k!}\nonumber\\
&\times\left[\int_{\beta}^{\infty} x^{m-n+k}
e^{-\frac{(x+\alpha)^2}{2}}dx\right.\nonumber\\
&\left.-(-1)^k\int_{\beta}^{\infty}
x^{m-n+k}e^{-\frac{(x-\alpha)^2}{2}}dx\right].
\end{align}
The calculation of the integral difference in
\eqref{eq:Integral_difference} can be effectively facilitated by the
following definition
\begin{align}\label{eq:Sotos_integral1}
\mathcal{I}_L^{k}(\alpha,\beta)={}&\int_{\beta}^{\infty} x^L
e^{-\frac{(x+\alpha)^2}{2}} dx-(-1)^k \int_{\beta}^{\infty}
x^Le^{-\frac{(x-\alpha)^2}{2}}dx
\end{align}
where $L=m-n+k$. Since we examine the case when $M \geq N$ or
equivalently $m \geq n$, it follows that in the above expression the
exponent $L$ is a non-negative integer. Therefore, using
\cite[eq.~(1.3.3.18)]{Prudnikov_v1_book}, \eqref{eq:Sotos_integral1}
obtains the form
\begin{align}\label{eq:Sotos_integral2}
\mathcal{I}_L^{k}(\alpha,\beta)={}&\sum_{l=0}^L \binom {L}{l}
\alpha^{L-l} \left[(-1)^{L-l} \int_{\beta+\alpha}^{\infty} x^l
e^{-\frac{x^2}{2}} dx\right.\nonumber\\
&\left.-(-1)^k \int_{\beta-\alpha}^{\infty}x^l e^{-\frac{x^2}{2}} dx
\right].
\end{align}
The two integrals involved in \eqref{eq:Sotos_integral2} can be
considered as special cases of the more general one
\begin{align*}
\mathcal{I}_b^l=\int_{b}^{\infty} x^l e^{-\frac{x^2}{2}}dx,
\quad b \in\mathds{R}
\end{align*}
which for the case of non-negative values of $b$ can be calculated
from \cite[eq.~(3.381.3)]{Gradshteyn_book} as
\begin{align}\label{eq:Sotos_integral3}
\mathcal{I}_b^l=2^{\frac{l-1}{2}}\Gamma{\left(\frac{l+1}{2},\frac{b^2}{2}\right)},
\quad b\geq 0
\end{align}
while for negative values of $b$, \cite[eqs.~(3.381.1),
(3.381.4)]{Gradshteyn_book} can be combined to yield
\begin{align}\label{eq:Sotos_integral4}
\mathcal{I}_b^l&=2^{\frac{l-1}{2}}\left[\Gamma{\left(\frac{l+1}{2}\right)}+(-1)^l
\gamma{\left(\frac{l+1}{2},\frac{b^2}{2}\right)}\right],
\quad b<0.
\end{align}
Therefore, a single expression for the integral $\mathcal{I}_b^l$
for any real value of $b$ can be derived, by merging
\eqref{eq:Sotos_integral3} and \eqref{eq:Sotos_integral4} with the
help of \cite[eq.~(8.356.3)]{Gradshteyn_book}, in order to satisfy
\begin{align*}
\mathcal{I}_b^l=2^{\frac{l-1}{2}}\left[\Gamma\left(\frac{l+1}{2}\right)-
[\sgn(b)]^{l+1}\gamma\left(\frac{l+1}{2},
\frac{b^2}{2}\right)\right].
\end{align*}
Thus, \eqref{eq:Sotos_integral1} is equivalent to
\begin{align*}
\mathcal{I}_L^{k}(\alpha,\beta)={}&\sum_{l=0}^L \binom {L}{l}
\alpha^{L-l}\left[(-1)^{L-l}\mathcal{I}_{\beta+\alpha}^l-(-1)^k\mathcal{I}_{\beta-\alpha}^l\right]\\
={}&(-1)^{k+1}\sum_{l=0}^L\binom{L}{l}2^{\frac{l-1}{2}}\alpha^{L-l}\left[\Gamma\left(\frac{l+1}{2}\right)\right.\\
&+(-1)^{L-l-k-1}\Gamma{\left(\frac{l+1}{2},\frac{(\beta+\alpha)^2}{2}\right)}\\
&\left.-[\sgn(\beta-\alpha)]^{l+1}\gamma\left(\frac{l+1}{2},\frac{(\beta-\alpha)^2}{2}\right)\right]
\end{align*}
which, after the substitution $L=m-n+k$, yields
\eqref{eq:Sotos_integral0}, thus completing the proof.
\end{IEEEproof}
\begin{corollary}[Closed-form for the normalized Nuttall Q]\label{corollary:norm_Nuttall_exact}
The normalized Nuttall \emph{Q}-function,
$\mathcal{Q}_{M,N}(\alpha,\beta)$, when $m=M+0.5 \in \mathds{N}$,
$n=N+0.5\in\mathds{N}$ and $M\geq N$, can be evaluated for
$\alpha>0$, $\beta\geq 0$ by the following closed-form expression:
\begin{align*}
\mathcal{Q}_{M,N}(\alpha,\beta)={}&\frac{(-1)^n
2^{-n+\frac{1}{2}}}{\sqrt{\pi}\alpha^{2n-1}}\\
&\times\sum_{k=0}^{n-1}\frac{(n-k)_{n-1}
(2\alpha)^k}{k!}\mathcal{I}_{m,n}^{k}(\alpha,\beta)
\end{align*}
where the term $\mathcal{I}_{m,n}^{k}(\alpha,\beta)$ is given by
\eqref{eq:Sotos_integral0}.
\end{corollary}
\begin{IEEEproof}
The proof follows immediately from
\eqref{eq:norm_Nuttall_definition} and
Theorem~\ref{theorem:st_Nuttall_exact}.
\end{IEEEproof}

\subsection{Lower and Upper Bounds}\label{subsec:Nuttall_bounds}

In this section, novel lower and upper bounds for the normalized and
standard Nuttall \emph{Q}-functions are proposed.
\begin{lemma}\label{lemma:gen_inc_gamma_ratio_monotonicity}
The function $\mathcal{G}_{s}(r,x)$, defined by
\begin{align}\label{eq:gen_inc_gamma_ratio}
\mathcal{G}_{s}(r,x)\triangleq\frac{\Gamma(r+s,x)}{\Gamma(r)}, \quad
r,x\in\mathds{R}^+
\end{align}
is strictly increasing with respect to $r$ for all
$s\in\mathds{R}^+_0$.
\end{lemma}
\begin{IEEEproof}
By multiplying both the numerator and denominator of
\eqref{eq:gen_inc_gamma_ratio} by the upper incomplete gamma
function, $\Gamma(r,x)$, we obtain
\begin{align*}
\mathcal{G}_{s}(r,x)=\frac{\Gamma(r+s,x)}{\Gamma(r,x)}\mathcal{G}_{0}(r,x)
\end{align*}
where from \eqref{eq:gen_inc_gamma_ratio} one can observe that the
term $\mathcal{G}_{0}(r,x)$ is the complement of the regularized
lower incomplete gamma function $P(r,x)$ with respect to unity,
defined in \cite[eq.~(6.5.1)]{Abramowitz_book} by
$P(r,x)=\frac{\gamma(r,x)}{\Gamma(r)}$ for all $r>0$ and
$x\in\mathds{R}$. Fortunately, $P(r,x)$ for $r,x>0$ is equal to the
cumulative distribution function (CDF) of the standard gamma
distribution $\verb"Gamma"(r,1)$, which is strictly decreasing with
respect to the shape parameter $r$. Additionally, this important result has
also been proved analytically in \cite[eq.~(59)]{Tricomi}, thus implying that
$\mathcal{G}_{0}(r,x)$ is strictly increasing with respect to $r>0$
for all $x>0$. Furthermore, in \cite{Qi_gamma_inequalities}, it has
been demonstrated that the function
\begin{align}\label{eq:Qi_inequality}
\mathcal{R}(p,q,x)=\left[\frac{\Gamma(p,x)}{\Gamma(q,x)}\right]^{\frac{1}{p-q}},
\quad p>q>0,x>0
\end{align}
is increasing with respect to $q$. By substituting $p=r+s$ and $q=r$
into \eqref{eq:Qi_inequality}, we realize that the ratio
$\Gamma(r+s,x)/\Gamma(r,x)$ is increasing with respect to $r$ for
$s>0$, while it remains constant for the trivial case of $s=0$.
Therefore, it increases with $r>0$ for all $x>0,s\geq 0$, and the
proof is complete.
\end{IEEEproof}

\begin{figure}[!t] \centering
\psfrag{xtitle1}[t][b][1]{\footnotesize $M+N$}%
\psfrag{ytitle1}[][][1]{\footnotesize $\mathcal{Q}_{M,N}(\alpha,\beta)$}%
\psfrag{xtitle2}[t][b][1]{\footnotesize $M+N$}%
\psfrag{ytitle2}[][][1]{\footnotesize $\mathcal{Q}_{M,N}(\alpha,\beta)$}%
\psfrag{vlonglegend1.1}[l][l][1]{\scriptsize $\alpha$=$7.5$, $\beta$=$6.5$}%
\psfrag{vlonglegend1.2}[l][l][1]{\scriptsize $\alpha$=$5.5$, $\beta$=$5.5$}%
\psfrag{vlonglegend1.3}[l][l][1]{\scriptsize $\alpha$=$0.5$, $\beta$=$3.5$}%
\psfrag{vlonglegend2.1}[l][l][1]{\scriptsize $\alpha$=$3.5$, $\beta$=$1.5$}%
\psfrag{vlonglegend2.2}[l][l][1]{\scriptsize $\alpha$=$2.0$, $\beta$=$2.0$}%
\psfrag{vlonglegend2.3}[l][l][1]{\scriptsize $\alpha$=$1.5$, $\beta$=$3.5$}%
\psfrag{x1.1}[][][1]{\footnotesize $4$}%
\psfrag{x1.2}[][][1]{\footnotesize $6$}%
\psfrag{x1.3}[][][1]{\footnotesize $8$}%
\psfrag{x1.4}[][][1]{\footnotesize $10$}%
\psfrag{x1.5}[][][1]{\footnotesize $12$}%
\psfrag{x1.6}[][][1]{\footnotesize $14$}%
\psfrag{x1.7}[][][1]{\footnotesize $16$}%
\psfrag{x1.8}[][][1]{\footnotesize $18$}%
\psfrag{y1.1}[r][r][1]{\footnotesize $0$}%
\psfrag{y1.2}[r][r][1]{\footnotesize $0.1$}%
\psfrag{y1.3}[r][r][1]{\footnotesize $0.2$}%
\psfrag{y1.4}[r][r][1]{\footnotesize $0.3$}%
\psfrag{y1.5}[r][r][1]{\footnotesize $0.4$}%
\psfrag{y1.6}[r][r][1]{\footnotesize $0.5$}%
\psfrag{y1.7}[r][r][1]{\footnotesize $0.6$}%
\psfrag{y1.8}[r][r][1]{\footnotesize $0.7$}%
\psfrag{y1.9}[r][r][1]{\footnotesize $0.8$}%
\psfrag{y1.10}[r][r][1]{\footnotesize $0.9$}%
\psfrag{y1.11}[r][r][1]{\footnotesize $1.0$}%
\psfrag{x2.1}[][][1]{\footnotesize $4$}%
\psfrag{x2.2}[][][1]{\footnotesize $6$}%
\psfrag{x2.3}[][][1]{\footnotesize $8$}%
\psfrag{x2.4}[][][1]{\footnotesize $10$}%
\psfrag{x2.5}[][][1]{\footnotesize $12$}%
\psfrag{x2.6}[][][1]{\footnotesize $14$}%
\psfrag{x2.7}[][][1]{\footnotesize $16$}%
\psfrag{x2.8}[][][1]{\footnotesize $18$}%
\psfrag{y2.1}[r][r][1]{\footnotesize $0$}%
\psfrag{y2.2}[r][r][1]{\footnotesize $1.0$}%
\psfrag{y2.3}[r][r][1]{\footnotesize $2.0$}%
\psfrag{y2.4}[r][r][1]{\footnotesize $3.0$}%
\psfrag{y2.5}[r][r][1]{\footnotesize $4.0$}%
\psfrag{y2.6}[r][r][1]{\footnotesize $5.0$}%
\psfrag{y2.7}[r][r][1]{\footnotesize $6.0$}%
\subfigure[$\mathcal{Q}_{M,N}(\alpha,\beta)$ versus $M+N$ for
$M-N=1$.] {\includegraphics[width=\linewidth,trim=10 0 35
10,clip=true]{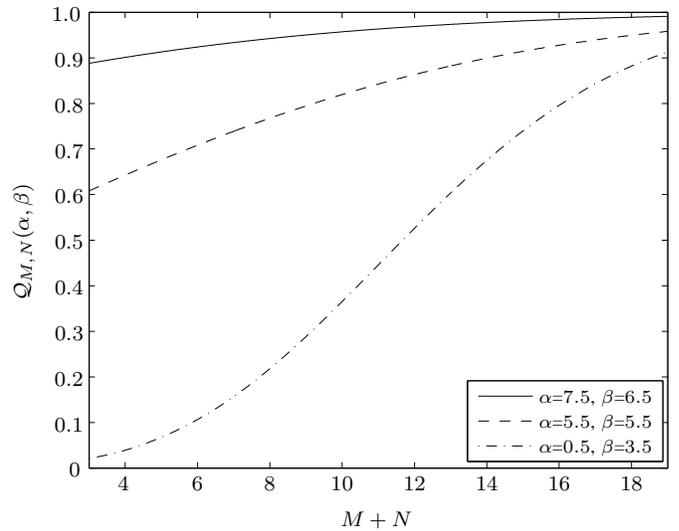}\label{fig:kapin1subfig1}}
\newline
\subfigure[$\mathcal{Q}_{M,N}(\alpha,\beta)$ versus $M+N$ for
$M-N=2$.] {\includegraphics[width=\linewidth,trim=10 0 35
10,clip=true]{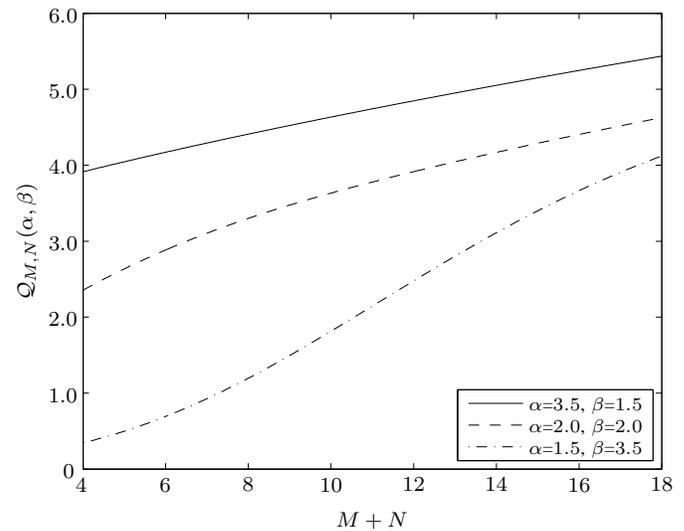}\label{fig:kapin1subfig2}}
\caption{Monotonicity of $\mathcal{Q}_{M,N}(\alpha,\beta)$ with
respect to the sum $M+N$ for several real values of
$\alpha,\beta$.}\label{fig:kapin1}
\end{figure}

The outcome of Lemma~\ref{lemma:gen_inc_gamma_ratio_monotonicity}
will be utilized for the establishment of the next theorem,
concerning the monotonicity property of the normalized Nuttall
\emph{Q}-function.
\begin{theorem}[Monotonicity of the normalized Nuttall Q]
\label{theorem:norm_Nuttall_monotonicity} The normalized Nuttall
\emph{Q}-function, $\mathcal{Q}_{M,N}(\alpha,\beta)$, where $M>0$,
$N>-1$ and $\alpha,\beta>0$, is strictly increasing with respect to
the sum $M+N$, under the requirement of constant difference $M-N
\geq 1$.
\end{theorem}
\begin{IEEEproof}
Combining \eqref{eq:st_Nuttall_definition},
\eqref{eq:norm_Nuttall_definition} and using the series
representation of the modified Bessel function of the first kind in
terms of the gamma function \cite[eq.~(8.445)]{Gradshteyn_book}, we
obtain
\begin{align}\label{eq:norm_Nuttall1}
\mathcal{Q}_{M,N}(\alpha,\beta)={}&e^{-\frac{\alpha^2}{2}}\sum_{k=0}^{\infty}\frac{\alpha^{2k}}
{k!\Gamma(k+N+1)2^{2k+N}}\nonumber\\
&\times\int_{\beta}^{\infty}x^{2k+M+N}e^{-\frac{x^2}{2}}dx
\end{align}
where we have interchanged the order of integration and summation,
since all integrand quantities of the normalized Nuttall
\emph{Q}-function are Riemann integrable on $[\beta,\infty)$.
Additionally, the integral in \eqref{eq:norm_Nuttall1} is the case
of \eqref{eq:Sotos_integral3}, thus yielding
\begin{align*}
\int_{\beta}^{\infty}x^{2k+M+N}e^{-\frac{x^2}{2}}dx={}&2^{k+\frac{M+N-1}{2}}\\
&\times\Gamma\left(k+\frac{M+N+1}{2},\frac{\beta^2}{2}\right).
\end{align*}
Therefore, \eqref{eq:norm_Nuttall1} reads
\begin{align}\label{eq:norm_Nuttall2}
\mathcal{Q}_{M,N}(\alpha,\beta)=e^{-\frac{\alpha^2}{2}}\sum_{k=0}^{\infty}\frac{\alpha^{2k}}
{2^{k+\frac{N-M+1}{2}}k!}\frac{\Gamma\left(k+\frac{M+N+1}{2},\frac{\beta^2}{2}\right)}{\Gamma(k+N+1)}.
\end{align}
Introducing the variables $v=M+N$ and $c=M-N$ and taking the partial
derivative of both sides of \eqref{eq:norm_Nuttall2} with respect to
$v$, we can easily obtain
\begin{align}\label{eq:norm_Nuttall_partial1}
\frac{\partial}{\partial{v}}{\mathcal{Q}_{\frac{v+c}{2},\frac{v-c}{2}}(\alpha,\beta)}={}&e^{-\frac{\alpha^2}{2}}
\sum_{k=0}^{\infty}\frac{\alpha^{2k}}{2^{k+\frac{3-c}{2}}k!}\nonumber\\
&\times\frac{\partial}{\partial{u(v)}}\mathcal{G}_{\frac{c-1}{2}}\left(u(v),\frac{\beta^2}{2}\right)
\end{align}
where the function $u(v)=k+1+\frac{v-c}{2}$ has been employed for
notational convenience. We note here that, applying the Weierstrass
M-test \cite{Titchmarsh_functions}, the series in
\eqref{eq:norm_Nuttall_partial1} can be proved to converge
uniformly, thus enabling one to interchange the order of
differentiation and summation. Hence, recalling
Lemma~\ref{lemma:gen_inc_gamma_ratio_monotonicity} and the
requirement of $\alpha>0$, that follows from the definition of the
normalized Nuttall \emph{Q}-function, the proof is complete.
\end{IEEEproof}
In Figs.~\ref{fig:kapin1}\subref{fig:kapin1subfig1} and
\ref{fig:kapin1}\subref{fig:kapin1subfig2}, the normalized Nuttall
\emph{Q}-function has been plotted versus the sum $M+N$ for several
values of $\alpha,\beta$, considering $M-N=1$ and $M-N=2$,
respectively. However, we note here that
Theorem~\ref{theorem:norm_Nuttall_monotonicity} implies non-integer
differences $M-N$ as well.

For the interpretation of the next proposition we define the pair of
\textit{half-integer rounding operators} $\lfloor x\rfloor_{0.5}$
and $\lceil x\rceil_{0.5}$ that map a real $x$ to its nearest left
and right half-odd integer, respectively, according to the
relations\footnote{The defining equations of
\eqref{eq:half-integer_operators} can be easily verified to be valid
in any arbitrary segment $[n,n+1)$, where $n\in\mathds{Z}$.}
\begin{align}
\begin{aligned}
\lfloor x\rfloor_{0.5}=\lfloor x-0.5\rfloor+0.5\\
\lceil x\rceil_{0.5}=\lceil x+0.5\rceil-0.5
\end{aligned}\label{eq:half-integer_operators}
\end{align}
where $\lfloor x\rfloor$ and $\lceil x\rceil$ denote the integer
floor and ceiling functions. Additionally, we recall that if
$\delta_x\in [0,1)$ is the fractional part of $x$, then $\lfloor
x\rfloor=x-\delta_x$.
\begin{corollary}[Bounds on the normalized Nuttall Q]\label{corollary:norm_Nuttall_bounds}
The following inequalities can serve as lower and upper bounds on
the normalized Nuttall \emph{Q}-function,
$\mathcal{Q}_{M,N}(\alpha,\beta)$, where $\alpha,\beta>0$ and
$M,N>0.5$, for the case when $M\geq N+1$ and $\delta_M=\delta_N$
(i.e. $M-N\in\mathds{N}$):
\begin{align}
\begin{aligned}
&\mathcal{Q}_{M,N}(\alpha,\beta)\geq\mathcal{Q}_{\lfloor M\rfloor_{0.5},\lfloor N\rfloor_{0.5}}(\alpha,\beta)\\
&\mathcal{Q}_{M,N}(\alpha,\beta)\leq\mathcal{Q}_{\lceil
M\rceil_{0.5},\lceil N\rceil_{0.5}}(\alpha,\beta).
\end{aligned}\label{eq:norm_Nuttall_inequality1}
\end{align}
with the equalities above being valid only for the case of half-odd
integer values of $M,N$.
\end{corollary}
\begin{IEEEproof}
The proof follows immediately from
Theorem~\ref{theorem:norm_Nuttall_monotonicity}.
\end{IEEEproof}

For the calculation of the bounds in
\eqref{eq:norm_Nuttall_inequality1}, the quantities
$\mathcal{Q}_{\lfloor M\rfloor_{0.5},\lfloor
N\rfloor_{0.5}}(\alpha,\beta)$ and $\mathcal{Q}_{\lceil
M\rceil_{0.5},\lceil N\rceil_{0.5}}(\alpha,\beta)$ can be evaluated
exactly by utilizing the results of
Corollary~\ref{corollary:norm_Nuttall_exact}. Moreover, for the case
of $M,N\in\mathds{N}$, the proposed bounds obtain the simplified
form
\begin{align}
\begin{aligned}
&\mathcal{Q}_{M,N}(\alpha,\beta)>\mathcal{Q}_{M-0.5,N-0.5}(\alpha,\beta)\\
&\mathcal{Q}_{M,N}(\alpha,\beta)<\mathcal{Q}_{M+0.5,N+0.5}(\alpha,\beta).
\end{aligned}\label{eq:norm_Nuttall_inequality2}
\end{align}

\begin{figure}[!t] \centering
\psfrag{xtitle1}[t][b][1]{\footnotesize $\beta$}%
\psfrag{ytitle1}[][][1]{\footnotesize $\mathcal{Q}_{5,3}(\alpha,\beta)$}%
\psfrag{xtitle2}[t][b][1]{\footnotesize $\beta$}%
\psfrag{ytitle2}[][][1]{\footnotesize $\mathcal{Q}_{M,2.7}(3.5,\beta)$}%
\psfrag{velonglegend1.1}[l][l][1]{\scriptsize $M$=$5.0$, $N$=$3.0$}%
\psfrag{velonglegend1.2}[l][l][1]{\scriptsize $M$=$4.5$, $N$=$2.5$}%
\psfrag{velonglegend1.3}[l][l][1]{\scriptsize $M$=$5.5$, $N$=$3.5$}%
\psfrag{gtext1.1}[][][1]{\small $\alpha=0.5$}%
\psfrag{gtext1.2}[][][1]{\small $\alpha=4$}%
\psfrag{gtext1.3}[][][1]{\small $\alpha=6.5$}%
\psfrag{gtext2.1}[][][1]{\small $M=3.7$}%
\psfrag{gtext2.2}[][][1]{\small $M=4.7$}%
\psfrag{gtext2.3}[][][1]{\small $M=5.7$}%
\psfrag{x1.1}[][][1]{\footnotesize $0$}%
\psfrag{x1.2}[][][1]{\footnotesize $2$}%
\psfrag{x1.3}[][][1]{\footnotesize $4$}%
\psfrag{x1.4}[][][1]{\footnotesize $6$}%
\psfrag{x1.5}[][][1]{\footnotesize $8$}%
\psfrag{x1.6}[][][1]{\footnotesize $10$}%
\psfrag{x1.7}[][][1]{\footnotesize $12$}%
\psfrag{y1.1}[r][r][1]{\footnotesize $0$}%
\psfrag{y1.2}[r][r][1]{\footnotesize $1.0$}%
\psfrag{y1.3}[r][r][1]{\footnotesize $2.0$}%
\psfrag{y1.4}[r][r][1]{\footnotesize $3.0$}%
\psfrag{y1.5}[r][r][1]{\footnotesize $4.0$}%
\psfrag{y1.6}[r][r][1]{\footnotesize $5.0$}%
\psfrag{y1.7}[r][r][1]{\footnotesize $6.0$}%
\psfrag{y1.8}[r][r][1]{\footnotesize $7.0$}%
\psfrag{y1.9}[r][r][1]{\footnotesize $8.0$}%
\psfrag{x2.1}[][][1]{\footnotesize $3$}%
\psfrag{x2.2}[][][1]{\footnotesize $4$}%
\psfrag{x2.3}[][][1]{\footnotesize $5$}%
\psfrag{x2.4}[][][1]{\footnotesize $6$}%
\psfrag{x2.5}[][][1]{\footnotesize $7$}%
\psfrag{x2.6}[][][1]{\footnotesize $8$}%
\psfrag{y2.1}[r][r][1]{\footnotesize $10^{-2}$}%
\psfrag{y2.2}[r][r][1]{\footnotesize $10^{-1}$}%
\psfrag{y2.3}[r][r][1]{\footnotesize $10^{0}$}%
\psfrag{y2.4}[r][r][1]{\footnotesize $10^{1}$}%
\psfrag{y2.5}[r][r][1]{\footnotesize $10^{2}$}%
\subfigure[$\mathcal{Q}_{5,3}(\alpha,\beta)$ versus $\beta$ for
several values of $\alpha$.]
{\includegraphics[width=\linewidth,trim=10 0 35
10,clip=true]{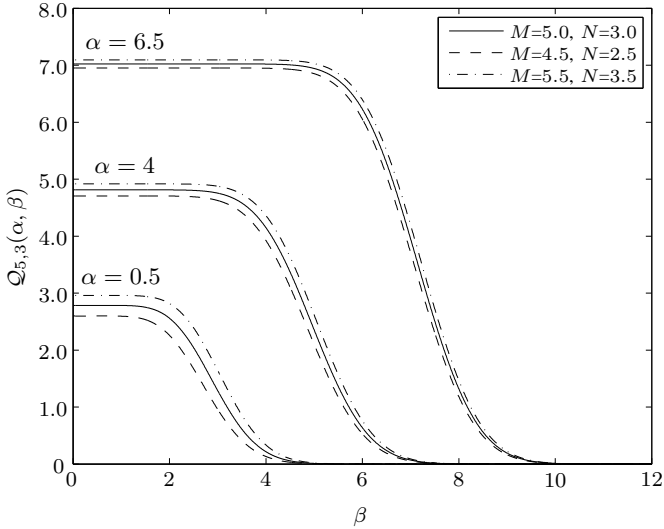}\label{fig:kapin2subfig1}}
\newline
\subfigure[$\mathcal{Q}_{M,2.7}(3.5,\beta)$ versus $\beta$ for
several values of $M$.] {\includegraphics[width=\linewidth,trim=10 0
35 10,clip=true]{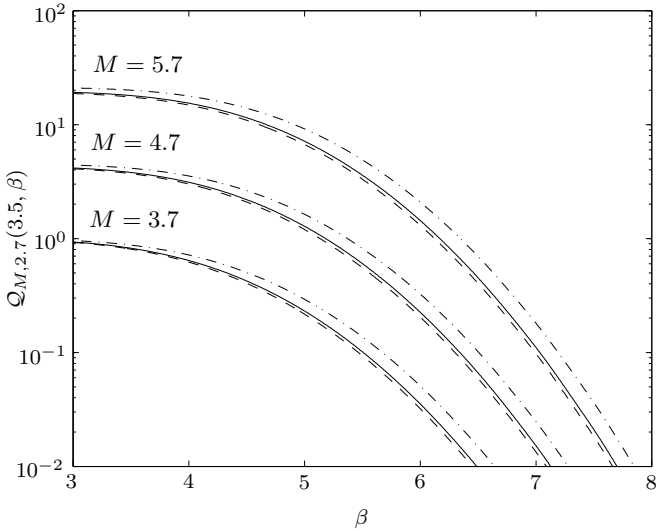}\label{fig:kapin2subfig2}}
\caption{Bounds of $\mathcal{Q}_{M,N}(\alpha,\beta)$ for
$M-N\in\mathds{N}$ and several values of
$\alpha,\beta$.}\label{fig:kapin2}
\end{figure}

In Figs.~\ref{fig:kapin2}\subref{fig:kapin2subfig1} and
\ref{fig:kapin2}\subref{fig:kapin2subfig2}, the normalized Nuttall
\emph{Q}-function along with its lower and upper bounds are depicted
versus $\beta$ for several values of $\alpha$ and $M$, respectively,
while the parameter $N$ is restricted according to the relation
$N=M-c$ with $c\in\mathds{N}$ taking values $c=2$ in
Fig.~\ref{fig:kapin2}\subref{fig:kapin2subfig1} and $c=1,2,3$ in
Fig.~\ref{fig:kapin2}\subref{fig:kapin2subfig2}. It is evident, that
the bounds proposed in \eqref{eq:norm_Nuttall_inequality1} are very
tight, especially the upper one for $\delta_M(=\delta_N)<0.5$ and
the lower one for $\delta_M(=\delta_N)>0.5$, the latter being the
case illustrated in Fig.~\ref{fig:kapin2}\subref{fig:kapin2subfig2}.

In order to obtain lower and upper bounds for the standard Nuttall
\emph{Q}-function, a similar procedure can be carried out. The next
theorem will be particularly useful for the fulfillment of such a
derivation.
\begin{theorem}[Monotonicity of the standard Nuttall Q]
\label{theorem:st_Nuttall_monotonicity} The standard Nuttall
\emph{Q}-function, $\emph{Q}_{M,N}(\alpha,\beta)$, where $M>0$,
$N>-1$ and $\alpha\geq 1$, $\beta>0$, is strictly increasing with
respect to the sum $M+N$, under the requirement of constant
difference $M-N \geq 1$.
\end{theorem}
\begin{IEEEproof}
In Theorem~\ref{theorem:norm_Nuttall_monotonicity}, it has been
proved that
\begin{align}\label{eq:norm_Nuttall_partial2}
\frac{\partial}{\partial{v}}{\mathcal{Q}_{\frac{v+c}{2},\frac{v-c}{2}}(\alpha,\beta)}>0
\end{align}
where we have substituted $v=M+N$ and $c=M-N$. From
\eqref{eq:norm_Nuttall_definition}, \eqref{eq:norm_Nuttall_partial2}
and after using the quotient rule for partial differentiation, we
obtain
\begin{align*}
\frac{\partial}{\partial{v}}{\emph{Q}_{\frac{v+c}{2},\frac{v-c}{2}}(\alpha,\beta)}>
\frac{\ln{\alpha}}{2}\emph{Q}_{\frac{v+c}{2},\frac{v-c}{2}}(\alpha,\beta).
\end{align*}
Since the Nuttall \emph{Q}-function is strictly positive, then for
$\alpha \geq 1$ it follows that
\begin{align*}
\frac{\partial}{\partial{v}}{\emph{Q}_{\frac{v+c}{2},\frac{v-c}{2}}(\alpha,\beta)}>0
\end{align*}
and the proof is complete.
\end{IEEEproof}
\begin{corollary}[Bounds on the standard Nuttall Q]\label{corollary:st_Nuttall_bounds}
The following inequalities can serve as lower and upper bounds on
the standard Nuttall \emph{Q}-function,
$\emph{Q}_{M,N}(\alpha,\beta)$, where $\alpha\geq 1$, $\beta>0$ and
$M,N>0.5$, for the case when $M\geq N+1$ and $\delta_M=\delta_N$
(i.e. $M-N\in\mathds{N}$):
\begin{align}
\begin{aligned}
&\emph{Q}_{M,N}(\alpha,\beta)\geq\emph{Q}_{\lfloor M\rfloor_{0.5},\lfloor N\rfloor_{0.5}}(\alpha,\beta)\\
&\emph{Q}_{M,N}(\alpha,\beta)\leq\emph{Q}_{\lceil
M\rceil_{0.5},\lceil N\rceil_{0.5}}(\alpha,\beta).
\end{aligned}\label{eq:st_Nuttall_inequality1}
\end{align}
with the equalities above being valid only for the case of half-odd
integer values of $M,N$.
\end{corollary}
\begin{IEEEproof}
The proof follows immediately from
Theorem~\ref{theorem:st_Nuttall_monotonicity}.
\end{IEEEproof}
Similarly to the case of the normalized Nuttall $\emph{Q}$, in the
calculation of the bounds from \eqref{eq:st_Nuttall_inequality1},
the quantities $\emph{Q}_{\lfloor M\rfloor_{0.5},\lfloor
N\rfloor_{0.5}}(\alpha,\beta)$ and $\emph{Q}_{\lceil
M\rceil_{0.5},\lceil N\rceil_{0.5}}(\alpha,\beta)$ can be evaluated
exactly from Theorem~\ref{theorem:st_Nuttall_exact}. Finally, for
$M,N\in\mathds{N}$, the standard Nuttall \emph{Q}-function can be
simply bounded by
\begin{align}
\begin{aligned}
\emph{Q}_{M,N}(\alpha,\beta)>\emph{Q}_{M-0.5,N-0.5}(\alpha,\beta)\\
\emph{Q}_{M,N}(\alpha,\beta)<\emph{Q}_{M+0.5,N+0.5}(\alpha,\beta)
\end{aligned}\label{eq:st_Nuttall_inequality2}
\end{align}
which constitutes the counterpart of
\eqref{eq:norm_Nuttall_inequality2} for the standard Nuttall
\emph{Q}-function.

\section{Monotonicity and Bounds for the Generalized Marcum \emph{Q}-Function}\label{sec:Marcum_monotonicity}

Recently, Li and Kam in \cite[eq.~(11)]{Li_conf_computing},
following a geometric approach, presented a novel closed-form
formula for the evaluation of $\emph{Q}_{M}(\alpha ,\beta )$, for
the case when $M$ is an odd multiple of $0.5$ and $\alpha>0$, $\beta
\geq 0$, given by
\begin{align}
\emph{Q}_{M}(\alpha,\beta)={}&\frac{1}{2}\text{erfc}\left(\frac{\beta+\alpha}{\sqrt
2}\right)+\frac{1}{2}\text{erfc}\left( \frac{\beta-\alpha}{\sqrt 2}
\right)\nonumber\\
&+\frac{1}{\alpha \sqrt{2\pi}} \sum_{k=0}^{M-1.5}
\frac{\beta^{2 k}}{2^k}\sum_{q=0}^{k} \frac{(-1)^q (2q)!}{(k-q)!q!}\nonumber\\
&\times\sum_{i=0}^{2q}\frac{1}{(\alpha\beta)^{2q-i}i!}\left[(-1)^i
e^{-\frac{(\beta-\alpha)^2}{2}}-e^{-\frac{(\beta+\alpha)^2}{2}}\right]\label{eq:Kam_Marcum_half}
\end{align}
where $\text{erfc}(z)=(2/\sqrt\pi)\int_z^\infty e^{-t^2}dt$ is the
complementary error function \cite[eq.~(7.1.2)]{Abramowitz_book}.
This representation involves only elementary functions and is
convenient for evaluation both numerically and analytically. For the
trivial case when $\alpha=0$, exact values of the generalized Marcum
\emph{Q}-function can be obtained from
\cite[eq.~(12)]{Li_conf_computing}
\begin{align}\label{eq:Kam_Marcum_zero}
\emph{Q}_{M}(0,\beta)={}&\text{erfc}\left(\frac{\beta}{\sqrt
2}\right)+\frac{e^{-\frac{\beta^2}{2}}}{\sqrt{2\pi}}\sum_{k=0}^{M-1.5}\frac{\beta^{2k+1}}{2^{k-1}}\nonumber\\
&\times\sum_{q=0}^{k}\frac{(-1)^q}{(k-q)!q!(2q+1)}.
\end{align}
Following an algebraic approach, an alternative more compact
closed-form expression, equivalent to \eqref{eq:Kam_Marcum_half},
can be derived, considering the next steps. Particularly, in
\cite[eq.~(10)]{Di_Blasio} it has been proved that the generalized
Marcum \emph{Q}-function of order $m-\mu$, with $m$ positive integer
and $0 \leq \mu < 1$ can be written in terms of the generalized
Marcum \emph{Q}-function of order $1-\mu$ as
\begin{align*}
\emph{Q}_{m-\mu}(\alpha,\beta)={}&e^{-\frac{{\alpha}^2+{\beta}^2}{2}}
\sum_{n=1}^{m-1} \left(\frac{\beta}{\alpha}\right)^{n-\mu}
I_{n-\mu}({\alpha}{\beta})\\
&+\emph{Q}_{1-\mu}(\alpha,\beta),
\quad {\alpha}\neq 0.
\end{align*}
By substituting $\mu=0.5$ in the above equation and noting that for
this case the modified Bessel function of the first kind can be
replaced by \eqref{eq:Sotos_mod_Bessel}, we obtain
\begin{align}
\emph{Q}_{m-0.5}(\alpha,\beta)={}&\alpha\sqrt{\frac{2}{\pi}}e^{-\frac{(\alpha+\beta)^2}{2}}
\sum_{n=1}^{m-1}(-2\alpha^2)^{-n}\nonumber\\
&\times\sum_{k=0}^{n-1}\frac{(n-k)_{n-1}}{k!}(2\alpha\beta)^k
\left[1-(-1)^k e^{2\alpha\beta}\right]\nonumber\\
&+\emph{Q}_{0.5}(\alpha,\beta),
\quad m\in\mathds{N}\label{eq:Revised_Marcum_half}
\end{align}
where once again $(m)_n$ denotes the Pochhammer's symbol and the
term $\emph{Q}_{0.5}(\alpha,\beta)$ can be derived from the
definition of the generalized Marcum \emph{Q}-function in
\eqref{eq:Marcum_definition}, by using
\cite[eq.~(10.2.14)]{Abramowitz_book} as follows
\begin{align*}
\emph{Q}_{0.5}(\alpha,\beta)=\sqrt{\frac{2}{\pi}}\int_{\beta}^{%
\infty}e^{-\frac{x^2+{\alpha}^2}{2}}\cosh{(ax)}dx.
\end{align*}
The above integral can be computed in closed-form as
\begin{align}
\emph{Q}_{0.5}(\alpha,\beta)={}&\frac{1}{2}
\text{erfc}\left(\frac{\beta+\alpha}{\sqrt 2}\right)+
\frac{1}{2}\text{erfc}\left(\frac{\beta-\alpha}{\sqrt 2}\right)\nonumber\\
={}&\emph{Q}(\beta+\alpha)+\emph{Q}(\beta-\alpha)\label{eq:Sotos_Marcum_half}
\end{align}
where \emph{Q} denotes the Gaussian \emph{Q}-function (or Gaussian
probability integral) \cite[eq.~(26.2.3)]{Abramowitz_book}, defined
by $\emph{Q}(z)=(1/\sqrt{2\pi})\int_z^\infty e^{-t^2/2}dt$. Using
\eqref{eq:Revised_Marcum_half} and \eqref{eq:Sotos_Marcum_half}, the
generalized Marcum \emph{Q}-function of half-odd integer order can
be computed for all $\alpha>0,\beta\geq 0$ from the expression
\begin{align}
\emph{Q}_{M}(\alpha,\beta)={}&\alpha\sqrt{\frac{2}{\pi}}e^{-\frac{(\alpha+\beta)^2}{2}}
\sum_{n=1}^{M-0.5}(-2\alpha^2)^{-n}\nonumber\\
&\times\sum_{k=0}^{n-1}\frac{(n-k)_{n-1}}{k!}(2\alpha\beta)^k\left[1-(-1)^k
e^{2\alpha\beta}\right]\nonumber\\
&+\emph{Q}(\beta+\alpha)+\emph{Q}(\beta-\alpha),\quad M+0.5\in\mathds{N}.\label{eq:Revised_Marcum_half_II}
\end{align}
We note here that a similar result to
\eqref{eq:Revised_Marcum_half_II} has been recently reported in the
literature \cite[eq.~(16)]{Sun+conf_tight}. In order to examine the
special case when $\alpha=0$, we first notice that from
\eqref{eq:Marcum-Nuttall_relation} and \eqref{eq:norm_Nuttall2} an
alternative expression---equivalent to
\cite[eq.~(26.4.25)]{Abramowitz_book}---for the generalized Marcum
\emph{Q}-function can be derived, written as
\begin{align}
\emph{Q}_{M}(\alpha,\beta)=e^{-\frac{\alpha^2}{2}}\sum_{k=0}^{\infty}\frac{\alpha^{2k}}
{2^{k}k!}\frac{\Gamma\left(k+M,\frac{\beta^2}{2}\right)}{\Gamma(k+M)},
\quad\alpha>0,\beta\geq 0\label{eq:Marcum_Series}
\end{align}
which for integer $M$ falls into the series expansion
\cite[eq.~(4)]{Dillard}. Since $\emph{Q}_{M}(\alpha,\beta)$ is a
continuous function of $\alpha$ for all $\beta\geq 0$ and $M>0$, the
above equation can be extended to be asymptotically valid for the
case when $\alpha=0$ as well, with the corresponding limiting value
given by
\begin{align}
\emph{Q}_{M}(0,\beta)=\frac{\Gamma\left(M,\frac{\beta^2}{2}\right)}{\Gamma(M)}.\label{eq:Sotos_Marcum_zero}
\end{align}
This last result also appears in
\cite[eq.~(4.71)]{Simon_Fading_channels_book}, where
$\emph{Q}_{M}(0,\beta)$ has been derived directly from
\eqref{eq:Marcum_definition} by applying the small argument form of
the modified Bessel function.

It has been proved in \cite{Li_conf_computing} that
\eqref{eq:Kam_Marcum_half}, \eqref{eq:Kam_Marcum_zero} along with
\eqref{eq:Kam_Marcum_inequality} can define tight upper and lower
bounds for the generalized Marcum \emph{Q}-function of integer
order. It seems apparent, that in order to derive bounds for
$\emph{Q}_{M}(\alpha,\beta )$ of real order $M$, a strict
inequality, involving the whole range of $M$, has to be established.
Such a generalization concept can be formalized through the
following theorem.
\begin{theorem}[Monotonicity of the generalized Marcum Q]
\label{theorem:Marcum_monotonicity} The generalized Marcum
\emph{Q}-function, $\emph{Q}_{M}(\alpha,\beta )$, is strictly
increasing with respect to its real order $M>0$ for all $\alpha\geq
0,\beta>0$.
\end{theorem}
\begin{IEEEproof}
Concerning the case when $\alpha=0$, we notice that
\eqref{eq:Sotos_Marcum_zero} can be rewritten as
\begin{align*}
\emph{Q}_{M}(0,\beta)=1-P\left(M,\frac{\beta^2}{2}\right).
\end{align*}
However, in \cite[eq.~(59)]{Tricomi} the regularized lower
incomplete gamma function $P(r,x)$ has been proved to decrease
monotonically with respect to $r>0$ for all $x>0$. Additionally, for
$\alpha>0$, \eqref{eq:Marcum-Nuttall_relation} implies that the
normalized Nuttall \emph{Q}-function with $N=M-1$ falls into the
generalized Marcum \emph{Q}-function of order $M$. Nevertheless,
according to Theorem~\ref{theorem:norm_Nuttall_monotonicity},
$\mathcal{Q}_{M,M-1}(\alpha,\beta)$ is strictly increasing with
respect to $2M-1$ for $M>0$, and the proof is complete.
\end{IEEEproof}

The result of Theorem~\ref{theorem:Marcum_monotonicity} has also
recently demonstrated by Sun and Baricz in
\cite{Baricz_inequalities}, where two totally different proofs were
given. The first one combines the series form of the generalized
Marcum \emph{Q}-function presented in \eqref{eq:Marcum_Series},
\eqref{eq:Sotos_Marcum_zero} together with the fact that the
regularized upper incomplete gamma function $Q(r,x)=1-P(r,x)$ is
strictly increasing with respect to $r>0$ for each $x>0$, originally
stated by Tricomi in \cite{Tricomi}. A slightly different analytical
proof to this can also be found in \cite[Th.~1]{Ghosh_monotonicity}.
The second proof exploits the interesting relationship between the
generalized Marcum \emph{Q}-function and the reliability function
(or CCDF) $R$ of a $\chi^2$ random variable with $2M$ DOF and
noncentrality parameter $\alpha$, namely the fact that if
$\beta\sim\chi_{2M,\alpha}^2$ then
$R(\beta)=\emph{Q}_{M}(\sqrt\alpha,\sqrt\beta)$. The interested
reader is referred to \cite[Th.~3.1]{Baricz_inequalities} for more
information.

\begin{figure}[!t]\centering
\psfrag{xtitle1}[][][1]{\footnotesize $\beta$}%
\psfrag{ytitle1}[][][1]{\footnotesize $\emph{Q}_4(\alpha,\beta)$}%
\psfrag{xtitle2}[][][1]{\footnotesize $\beta$}%
\psfrag{ytitle2}[][][1]{\footnotesize $\emph{Q}_M(2.5,\beta)$}%
\psfrag{legen1.1}[l][l][1]{\scriptsize $M$=$4.0$}%
\psfrag{legen1.2}[l][l][1]{\scriptsize $M$=$3.5$}%
\psfrag{legen1.3}[l][l][1]{\scriptsize $M$=$4.5$}%
\psfrag{gtext1.1}[][][1]{\small $\alpha=0.5$}%
\psfrag{gtext1.2}[][][1]{\small $\alpha=3.5$}%
\psfrag{gtext1.3}[][][1]{\small $\alpha=5.5$}%
\psfrag{gtext2.1}[][][1]{\small $M=2.7$}%
\psfrag{gtext2.2}[][][1]{\small $M=8.3$}%
\psfrag{x1.1}[][][1]{\footnotesize $1$}%
\psfrag{x1.2}[][][1]{\footnotesize $2$}%
\psfrag{x1.3}[][][1]{\footnotesize $3$}%
\psfrag{x1.4}[][][1]{\footnotesize $4$}%
\psfrag{x1.5}[][][1]{\footnotesize $5$}%
\psfrag{x1.6}[][][1]{\footnotesize $6$}%
\psfrag{x1.7}[][][1]{\footnotesize $7$}%
\psfrag{x1.8}[][][1]{\footnotesize $8$}%
\psfrag{x1.9}[][][1]{\footnotesize $9$}%
\psfrag{y1.1}[r][r][1]{\footnotesize $0$}%
\psfrag{y1.2}[r][r][1]{\footnotesize $0.1$}%
\psfrag{y1.3}[r][r][1]{\footnotesize $0.2$}%
\psfrag{y1.4}[r][r][1]{\footnotesize $0.3$}%
\psfrag{y1.5}[r][r][1]{\footnotesize $0.4$}%
\psfrag{y1.6}[r][r][1]{\footnotesize $0.5$}%
\psfrag{y1.7}[r][r][1]{\footnotesize $0.6$}%
\psfrag{y1.8}[r][r][1]{\footnotesize $0.7$}%
\psfrag{y1.9}[r][r][1]{\footnotesize $0.8$}%
\psfrag{y1.10}[r][r][1]{\footnotesize $0.9$}%
\psfrag{y1.11}[r][r][1]{\footnotesize $1.0$}%
\psfrag{x2.1}[][][1]{\footnotesize $1$}%
\psfrag{x2.2}[][][1]{\footnotesize $2$}%
\psfrag{x2.3}[][][1]{\footnotesize $3$}%
\psfrag{x2.4}[][][1]{\footnotesize $4$}%
\psfrag{x2.5}[][][1]{\footnotesize $5$}%
\psfrag{x2.6}[][][1]{\footnotesize $6$}%
\psfrag{x2.7}[][][1]{\footnotesize $7$}%
\psfrag{y2.1}[r][r][1]{\footnotesize $0$}%
\psfrag{y2.2}[r][r][1]{\footnotesize $0.1$}%
\psfrag{y2.3}[r][r][1]{\footnotesize $0.2$}%
\psfrag{y2.4}[r][r][1]{\footnotesize $0.3$}%
\psfrag{y2.5}[r][r][1]{\footnotesize $0.4$}%
\psfrag{y2.6}[r][r][1]{\footnotesize $0.5$}%
\psfrag{y2.7}[r][r][1]{\footnotesize $0.6$}%
\psfrag{y2.8}[r][r][1]{\footnotesize $0.7$}%
\psfrag{y2.9}[r][r][1]{\footnotesize $0.8$}%
\psfrag{y2.10}[r][r][1]{\footnotesize $0.9$}%
\psfrag{y2.11}[r][r][1]{\footnotesize $1.0$}%
\subfigure[$\emph{Q}_4(\alpha,\beta)$ versus $\beta$ for several
values of $\alpha$.] {\includegraphics[width=\linewidth,trim=10 0 35
10,clip=true]{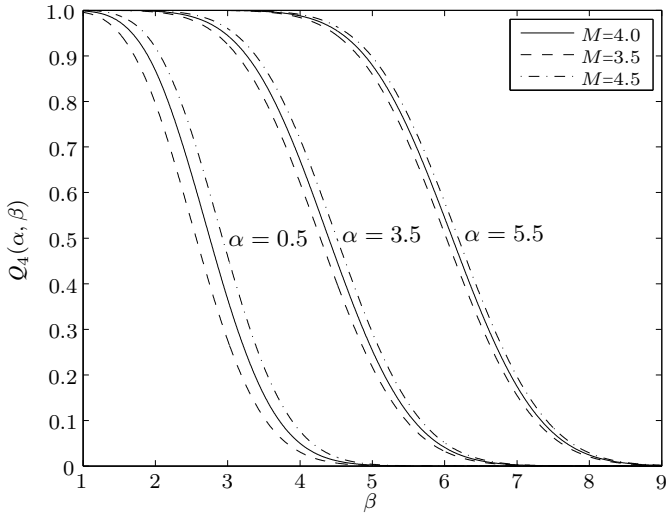}\label{fig:kapin3subfig1}}
\newline
\subfigure[$\emph{Q}_M(2.5,\beta)$ versus $\beta$ for several values
of $M$.] {\includegraphics[width=\linewidth,trim=10 0 35
10,clip=true]{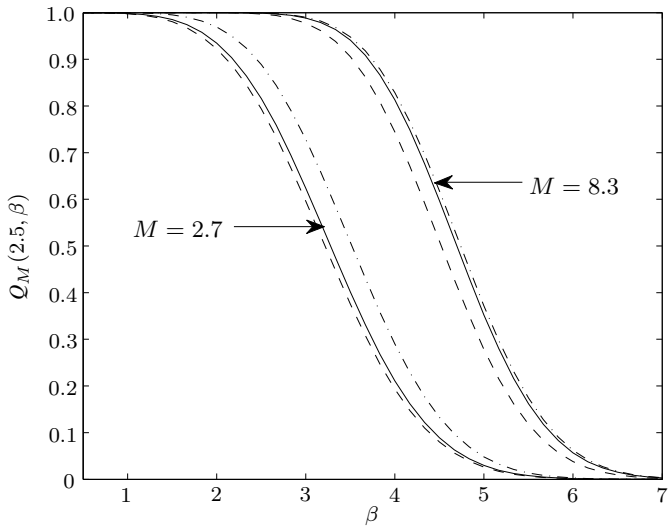}\label{fig:kapin3subfig2}} \caption{Bounds
of $\emph{Q}_M(\alpha,\beta)$ for several real values of
$\alpha,\beta$ and $M$. (In \subref{fig:kapin3subfig1} bounds
proposed by Li and Kam,
\cite{Li_conf_computing}).}\label{fig:kapin3}
\end{figure}

Recalling the relation between the normalized Nuttall and the
generalized Marcum \emph{Q}-functions, that is
\eqref{eq:Marcum-Nuttall_relation},
Fig.~\ref{fig:kapin1}\subref{fig:kapin1subfig1} verifies graphically
the results of Theorem~\ref{theorem:Marcum_monotonicity}, since it
actually depicts $\emph{Q}_{M}(\alpha,\beta )$ versus the term
$2M-1$.
\begin{corollary}[Bounds on the generalized Marcum Q]\label{corollary:st_Marcum_Bounds}
The following inequalities can serve as lower and upper bounds on
the generalized Marcum \emph{Q}-function
$\emph{Q}_{M}(\alpha,\beta)$ of real order $M>0.5$ for all
$\alpha\geq 0,\beta>0$.
\begin{align}
\emph{Q}_{\lfloor
M\rfloor_{0.5}}(\alpha,\beta)\leq\emph{Q}_M(\alpha,\beta)\leq\emph{Q}_{\lceil
M\rceil_{0.5}}(\alpha,\beta).\label{eq:Marcum_inequality1}
\end{align}
with the equalities above being valid only for the case of half-odd
integer values of $M$.
\end{corollary}
\begin{IEEEproof}
The proof follows immediately from
Theorem~\ref{theorem:Marcum_monotonicity}.
\end{IEEEproof}
In Corollary~\ref{corollary:st_Marcum_Bounds}, the quantities
$\emph{Q}_{\lfloor M\rfloor_{0.5}}(\alpha,\beta)$ and
$\emph{Q}_{\lceil M\rceil_{0.5}}(\alpha,\beta)$ can be evaluated
exactly either from \eqref{eq:Kam_Marcum_half},
\eqref{eq:Kam_Marcum_zero} or \eqref{eq:Sotos_Marcum_half},
\eqref{eq:Sotos_Marcum_zero}, while for $M\in\mathds{N}$
\eqref{eq:Marcum_inequality1} reduces to
\begin{align*}
\emph{Q}_{M-0.5}(\alpha,\beta)<\emph{Q}_{M}(\alpha,\beta)<
\emph{Q}_{M+0.5}(\alpha,\beta)
\end{align*}
which comes as a complement to the inequalities of
\eqref{eq:norm_Nuttall_inequality2} and
\eqref{eq:st_Nuttall_inequality2}. This last result was originally
demonstrated in \cite[eq.~(16)]{Li_conf_computing}, where the
authors following a geometric approach proposed tight lower and
upper bounds for the generalized Marcum \emph{Q}-function of integer
order $M$, which have been proved to outperform other existing ones.
This can be easily verified from
Fig.~\ref{fig:kapin3}\subref{fig:kapin3subfig1}, where
$\emph{Q}_4(\alpha,\beta)$ has been plotted versus $\beta$ for
several values of $\alpha$. Therefore, for the case of real $M$, one
can expect even further enhancement in the strictness of either the
lower bound (for $\delta_M>0.5$) or the upper one (for
$\delta_M<0.5$). This is clearly depicted in
Fig.~\ref{fig:kapin3}\subref{fig:kapin3subfig2}, where the curves
$\emph{Q}_{2.5}(2.5,\beta )$ and $\emph{Q}_{8.5}(2.5,\beta )$
constitute very tight lower and upper bounds of
$\emph{Q}_{2.7}(2.5,\beta )$ and $\emph{Q}_{8.3}(2.5,\beta )$,
respectively, for all range of $\beta$.

\section{Conclusion}\label{sec:Conclusion}

Applicable monotonicity criteria were established for the normalized
and standard Nuttall and the generalized Marcum \emph{Q}-functions.
Specifically, it was proved that the two Nuttall \emph{Q}-functions
are strictly increasing with respect to the real sum $M+N$ for the
case when $M\geq N+1$, while the generalized Marcum
\emph{Q}-function increases monotonically with respect to its real
order $M$. Additionally, novel closed-form expressions for both
types of the Nuttall \emph{Q}-function were given for the case when
$M,N$ are odd multiples of $0.5$ and $M \geq N$. Regarding the
generalized Marcum \emph{Q}-function of half-odd integer order, an
alternative more compact closed-form expression, equivalent to the
already existing one, was derived. By exploiting these results,
novel lower and upper bounds were proposed for the Nuttall
\emph{Q}-functions when $M\geq N+1$, while the recently proposed
bounds for the generalized Marcum \emph{Q}-function of integer $M$,
were appropriately utilized in order to extend their validity over
real values of $M$.



\begin{IEEEbiographynophoto}
{Vasilios M. Kapinas} (S'07--M'09) was born in Thessaloniki, Greece,
in May 1976. He received the diploma degree in electrical and
computer engineering from Aristotle University of Thessaloniki,
Greece, in 2000. Since 2005, he has been working toward the Ph.D.
degree in telecommunications engineering.

His current research interests include wireless communication theory
and digital communications over fading channels, giving special
focus to space-time block coding techniques.
\end{IEEEbiographynophoto}

\begin{IEEEbiographynophoto}
{Sotirios K. Mihos} was born in Thessaloniki, Greece, in April 1984.
He is an undergraduate student at the Aristotle University of
Thessaloniki, Greece, where he is working toward the diploma degree
in electrical and computer engineering.

His research interests span a wide range of subject areas including
computer science, electronics and automatic control, with a special
focus on their relationship to pure mathematics.
\end{IEEEbiographynophoto}

\begin{IEEEbiographynophoto}
{George K. Karagiannidis} (M'97--SM'04) was born in Pithagorion,
Samos Island, Greece. He received the University and Ph.D. degrees
in electrical engineering from the University of Patras, Patras,
Greece, in 1987 and 1999, respectively. From 2000 to 2004, he was a
Senior Researcher at the Institute for Space Applications and Remote
Sensing, National Observatory of Athens, Greece. In June 2004, he
joined Aristotle University of Thessaloniki, Thessaloniki, Greece,
where he is currently an Assistant Professor in the Electrical and
Computer Engineering Department. His current research interests
include wireless communication theory, digital communications over
fading channels, cooperative diversity systems, cognitive radio,
satellite communications, and wireless optical communications.

He is the author or coauthor of more than 80 technical papers
published in scientific journals and presented at international
conferences. He is also a coauthor of two chapters in books and a
coauthor of the Greek edition of a book on mobile communications. He
serves on the editorial board of the \textsc{EURASIP Journal on
Wireless Communications and Networking}.

Dr. Karagiannidis has been a member of Technical Program Committees
for several IEEE conferences. He is a member of the editorial boards
of the \textsc{IEEE Transactions on Communications} and the
\textsc{IEEE Communications Letters}. He is co-recipient of the Best
Paper Award of the Wireless Communications Symposium (WCS) in IEEE
International Conference on Communications (ICC' 07), Glasgow, U.K.,
June 2007. He is a full member of Sigma Xi.
\end{IEEEbiographynophoto}

\end{document}